\documentclass[11pt]{article}
\usepackage{moriond,epsfig}
\usepackage{floatflt}

\def\Journal#1#2#3#4{{#1} {\bf #2}, #3 (#4)}

\def\PRD{{\em Phys. Rev.} D}

\def\be{\begin{equation}}
\def\ee{\end{equation}}
\def\bea{\begin{eqnarray}}
\def\eea{\end{eqnarray}}

\begin{document}
\vspace*{4cm}
\title{GMSB SUSY MODELS WITH NON POINTING PHOTONS SIGNATURES \\ IN ATLAS AT THE LHC}

\author{ D. PRIEUR }

\address{Laboratoire d'Annecy-le-vieux de Physique des Particules,\\ 9 chemin de Bellevue, BP 110,
74941 Annecy-le-vieux, France}

\maketitle\abstracts{ The reconstruction of non pointing photons is
a key feature for studying gauge mediated supersymmetry breaking
(GMSB) models at the LHC. In this article the angular resolution of
the ATLAS electromagnetic calorimeter is characterized from a
detailed simulation of the detector. Resulting performances are used
to reconstruct GMSB events with a fast simulation program, taking
into account reconstruction effects. Finally, the sensitivity to
extract the sparticles masses and the lightest neutralino lifetime
is estimated. }

\section{Gauge mediated supersymmetry breaking models and non pointing photons}
The origin of the supersymmetry (SUSY) breaking and its mediation to
the MSSM sector are key features of SUSY models. In gauge mediated
SUSY breaking (GMSB) models\cite{Giudice:1998bp}, the breaking of
SUSY takes place in a hidden sector at a high energy scale
$\sqrt{F_0}$. Contrary to SUGRA type models, SUSY breaking is not
generated at the Planck scale but at a much lower energy scale. This
breaking is then transmitted to the MSSM sector through chiral
superfields belonging to an intermediate messenger sector at energy
scale $M_{mess}$. The coupling between the messenger and the MSSM
sector is made through classical $SU(3)_c\otimes SU(2)_L\otimes
U(1)_Y$ gauge interactions. Gravitational interactions are still
present but their contributions are small. Since the gravitino
$\widetilde{G}$ gets its mass only through gravitational
interaction, it is the lightest SUSY particle (LSP). We assume here
that the R-parity is conserved so that all heavier SUSY particles
will produce decay chains leading to the production of gravitinos.
The minimal GMSB model is driven by six arbitrary parameters.
Depending on these parameters, the next to lighest SUSY particle
(NLSP) can either be the lighest neutralino ($\widetilde{\chi}^0_1$)
or a right handed slepton ($\widetilde{l}_R$). One of the feature of
GMSB models is that the NLSP lifetime $c\tau$ may be macroscopic and
can vary from micrometers to kilometers. The value of $c\tau$ is
linked to $m_{NLSP}$ and to $\sqrt{F_0}$ through the
relation\cite{Ambrosanio:1996jn}
\begin{equation}
c\tau = \frac{1}{k_\gamma} \left( \frac{100~GeV}{m_{NLSP}} \right)^5
\left( \frac{\sqrt{F_0}}{100~TeV}\right)^4 \times 10^{-2}~cm
\label{eqn:ctau}
\end{equation}
where $k_\gamma\equiv |N_{11}cos\theta_W+N_{12}sin\theta_W|$ with
$\theta_W$ the Weinberg angle and $N_{ij}$ the mixing angles of the
neutralinos.

We consider here that the NLSP is the lightest neutralino, so that
the dominant decay mode is $\widetilde{\chi}^0_1 \rightarrow \gamma
\widetilde{G}$. If the NLSP has an intermediate lifetime, its decay
products will emerge away from the primary vertex: missing energy
and non pointing photons will be the signature for such a decay.
Measuring the NLSP lifetime provide a way to access $\sqrt{F_0}$,
the fundamental supersymmetric breaking scale. This can be done by
reconstructing the decay vertex of the neutralino from the photon
direction and its time of arrival.

In the following, the reconstruction of the photon direction using
the electromagnetic calorimeter is explained and the polar angular
resolution of non pointing photon is caracterized from a detailed
simulation of the ATLAS detector. Resulting performances are used to
reconstruct the specific decay chain $\widetilde{l} \rightarrow
\widetilde{\chi}^0_1 l \rightarrow \widetilde{G}l\gamma$ with a fast
simulation program, taking into account reconstruction effects, to
determine the mass and the lifetime of $\widetilde{\chi}^0_1$ and
finally the sensitivity to $\sqrt{F_0}$.

\section{ATLAS electromagnetic calorimeter}

\subsection{Description}
\begin{floatingfigure}[r]{6.5cm}
\begin{center}
    \includegraphics[width=6.5cm]{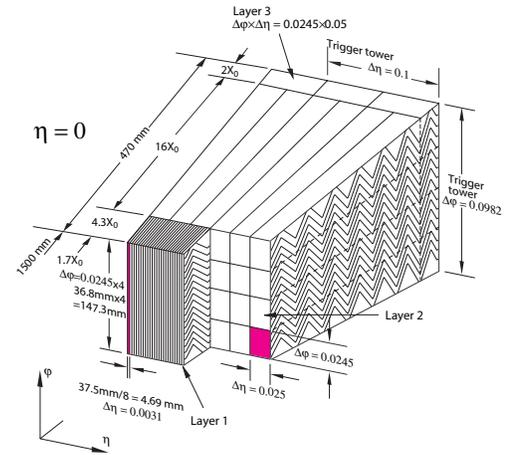}
    \caption{Readout cells granularity of the ATLAS electromagnetic calorimeter. }
    \label{fig:granularity}
\end{center}
\end{floatingfigure}

The ATLAS electromagnetic calorimeter\cite{CaloPerf} is a projective
calorimeter with a fine granularity to perform precision
measurements of the shower position. It is longitudinally divided
into three layers (figure \ref{fig:granularity}). The first layer is
longitudinally segmented along $\eta$ into very thin cells of
$\Delta \eta=0.003125$, leading to a resolution on $\eta$ position
with pointing photons of $0.30\times 10^{-3}$. The second layer has
a wider $\eta$ granularity of $\Delta \eta=0.025$ and it is designed
to contain most of the shower energy. It has a resolution on $\eta$
position of $0.83\times 10^{-3}$. By combining the measurement of
the $\eta$ position in the first two layers ($\eta_1$ and $\eta_2$),
it is possible to determine the shower direction along $\eta$.
Having a parametrization of the shower depth\cite{CaloPerf} for each
layers ($R_1(\eta_1)$ and $R_2(\eta_2)$), the shower polar direction
$\eta_{pointing}$ is reconstructed using the relation:
\begin{equation}
sinh(\eta_{pointing}) =
\frac{R_2(\eta_2)sinh(\eta_2)-R_1(\eta_1)sinh(\eta_1)}{R_2(\eta_2)-R_1(\eta_1)}
\end{equation}

It is important to notice that the coarse granularity of the layers
along the $\phi$ direction do not permit to reconstruct the $\phi$
direction of non pointing photons using only the electromagnetic
calorimeter.

\subsection{Polar angular resolution}

\begin{floatingfigure}[r]{8cm}
\begin{center}
    \includegraphics[width=8cm]{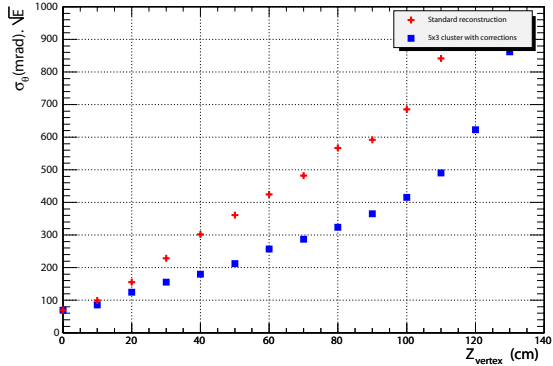}
    \caption{Polar angular resolution $\sigma_\theta$ for $\left| \eta \right| < 1.4$ before (red crosses) et after (blue squares) corrections, for non pointing photons generated
    with differents $Z_{vertex}$ position along the beam axis.}
    \label{fig:angres}
\end{center}
\end{floatingfigure}

The angular performances of the electromagnetic calorimeter were
studied with a detailed simulation\cite{Dice} of the ATLAS detector.
Several sets of $60$~GeV singles photons were generated at
differents positions along the beam axis with $\left| \eta
\right|<2.5$. The reconstruction of all events was done using ATLAS
standard reconstruction software. No electronic noise or pile-up
effects were added during the reconstruction. If the contributions
of the electronic noise should be small, influence of pile-up on the
angular resolution will have to be studied.

The angular resolution achieved using the standard reconstruction
with pointing photons is $\sigma_\theta=60$~mrad/$\sqrt{E(GeV)}$.
The results concerning the non pointing photons are shown in figure
\ref{fig:angres}. The performances are worsened as the position of
the photon generation vertex increase along the z axis. The
deterioration comes from several systematic biais at differents
level of the reconstruction algorithms. The S-shape effect is a
distortion of the reconstructed $\eta$ position due to the finite
cluster size. The S-shape corrections, that were tuned for pointing
photons are no more valid. The $3\times3$ cells cluster size used is
not sufficient to contain all the electromagnetic shower and some
energy leakage outside the cluster is possible. Finally, for large
deviation from pointing, the shower depth parametrization is no
longer valid.

To improve the resolution several changes have been made. The
cluster size has been extended to $5\times3$ cells and S-shape
corrections are not applied. An iterative algorithm has been
developed to correct for the systematic biais of the reconstructed
position in each layer. Results of this correction on the angular
resolution is presented on figure \ref{fig:angres}. The resolution
can be improved by $30$ to $40\%$ for photons coming from an
effective vertex up to $100$~cm along the beam axis. This polar
angular resolution has been parametrized and is used in the
following analysis.

\vspace{1cm}

\section{Study of $\widetilde{l} \rightarrow
\widetilde{\chi}^0_1 l \rightarrow \widetilde{G}l\gamma$ decay
channel}

From this point we consider the specific decay chain $\widetilde{l}
\rightarrow \widetilde{\chi}^0_1 l \rightarrow \widetilde{G}l\gamma$
leading to the production of non pointing photons. It is possible to
solve the gravitino momentum and the $\widetilde\chi^0_1$ decay
position for this cascade decay\cite{Kawagoe:2003jv} by knowing the
energy of the lepton, the energy and the time of arrival of the
photon, and by reconstructing the three angles between the lepton,
the photon and the gravitino. The analysis takes place at the GMSB
point G1 (table \ref{tab:pointG1}), with
$m_{\widetilde{\chi}^0_1}=117$~GeV, $m_{\widetilde{l}_R}=162$~GeV
and we consider a neutralino lifetime $c\tau$ between $10$~cm and
$2$~m. Two sets of $10^5$ and $10^6$ SUSY events were generated,
corresponding to one year of LHC at respectively low and nominal
luminosity. Final state particles were passed through the fast
simulation of the ATLAS detector and observables were smeared
according to realistic resolutions from test-beam data or detailed
simulation (table \ref{tab:resolutions}).

\begin{table}[h]
\begin{center}
\begin{tabular}{|ccccccc|}
\hline
Point & $\Lambda$ (TeV) & $M_{mess}$ (TeV) & $N_5$ & $tan\beta$ & $sgn(\mu )$ & $C_{grav}$ \\
\hline G1 & $90$ & $500$ & $1$ & $5.0$ & $+$ & - \\
\hline
\end{tabular}
\caption{GMSB model parameters at point G1.} \label{tab:pointG1}
\end{center}
\end{table}

\begin{table}[h]
\begin{center}
\begin{tabular}{|lcc|}
\hline
Observable & Sub-detector & Resolution \\
\hline
Energy & EM CAL. & $\frac{\delta E}{E} = \frac{10\%}{\sqrt{E}}\oplus\frac{245~MeV}{E}\oplus 0.7\%$ \\
Time & EM CAL. & $\sigma_t = 100~ps$ \\
Position & EM CAL. & $\sigma_\eta=\frac{0.004}{\sqrt{E(GeV)}}$, $\sigma_\phi=\frac{5~mrad}{\sqrt{E(GeV)}}$\\
Direction & EM CAL. & $\sigma_\theta$ \\
 & TRT & $\sigma_{\Delta\phi}=1~mrad$ \\
 \hline
\end{tabular}
\caption{Resolution applied to the reconstructed non pointing
photons observables.} \label{tab:resolutions}
\end{center}
\end{table}

\subsection{Reconstruction of sparticle masses}

In this part we consider a typical lifetime for the NLSP of
$100$~cm. To reconstruct $m_{\widetilde{\chi}^0_1}$ and
$m_{\widetilde{l}_R}$ we have to fully determine the direction of
the photon. Since the electromagnetic calorimeter can only
reconstruct the polar direction we require the photon to convert
inside the fiducial volume of the inner detector, in order to
measure $\phi$ with the TRT detector.

Standard pre-selection cuts\cite{Hinchliffe:1998ys} to limit
background contribution from standard model are applied. An
effective mass $M_{eff}$ is defined as the sum of the missing
transverse energy $E_T^{miss}$ and the transverse momentum of the
four hardest jets. Requiring $M_{eff}>400$~GeV,
$E_T^{miss}>0.1M_{eff}$ and at least two leptons and two photons for
each events has an efficiency of $55\%$. Converted non pointing
photons candidates are then selected by imposing $E_\gamma> 30$~GeV,
an non pointing angle $\alpha>0.2$~rad and an arrival time $\Delta
t_\gamma>1$~ns. Finally each selected non pointing photon is paired,
if possible, with an isolated lepton with $p_T>20~$GeV.
Lepton/photon pairs are used to solve the decay chain kinematic
relations and to determine sparticles masses. At GMSB point G1, for
$c\tau_{\widetilde{\chi}^0_1}=100$~cm, the mass resolution is
$\sigma_{\widetilde{\chi}^0_1}=1.7$~GeV and
$\sigma_{\widetilde{l}_R}=2.1$~GeV for $10^5$ generated SUSY events.

\subsection{Reconstruction of NLSP lifetime}
Once the sparticle masses are known, the entire decay chain can be
reconstructed. At this stage we no longer need to know the photon
$\phi$ direction and so can remove the requirement on conversion.
The photon $\phi$ direction and the position of the
$\widetilde{\chi}^0_1$ decay vertex are extracted by using a
minimization procedure. The same cuts as in the previous part are
applied, except the cut on the non pointing angle $\alpha$ which
becomes a cut on the polar direction $\theta$ of the photon.

\begin{figure}[tb]
\begin{center}
    \includegraphics[width=.6\textwidth]{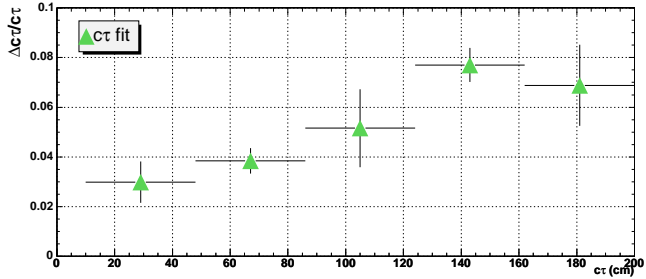}
    \caption{Sensitivity on the measure of the $\widetilde{\chi}^0_1$ lifetime $c\tau$.}
    \label{fig:ctausensitivity}
\end{center}
\end{figure}

To study the sensitivity of the reconstructed NLSP lifetime, we make
$c\tau_{\widetilde{\chi}^0_1}$ vary from $10$~cm up to $200$~cm. The
$\widetilde{\chi}^0_1$ lifetime is extracted by fitting an
exponential function to the proper time distribution of the
$\widetilde{\chi}^0_1$ in the laboratory frame. For the studied
$c\tau_{\widetilde{\chi}^0_1}$ range, the sensitivity $\Delta
c\tau/c\tau$ is found to vary from $3$ to $8\%$ (figure
\ref{fig:ctausensitivity}). For
$c\tau_{\widetilde{\chi}^0_1}=100$~cm, one can expect to reconstruct
the $\widetilde{\chi}^0_1$ mass with a $2\%$ precision and a $5\%$
precision on $c\tau_{\widetilde{\chi}^0_1}$. The sensitivity on the
fundamental SUSY breaking scale $\sqrt{F_0}$ and on the gravitino
mass $m_{\widetilde{G}}$ are respectively $4\%$ and $8\%$.
Statistical methods to extend the accessible $c\tau$ range are under
consideration: they require that systematics errors and detector
acceptance are carefully studied with a detailed simulation.



\section*{References}
\begin{thebibliography}{99}

\bibitem{Giudice:1998bp} G.F. Giudice and R. Rattazzi, \Journal{Phys.
Rept.}{322}{419}{1999}

\bibitem{Ambrosanio:1996jn} S. Ambrosanio {\it et al},
\Journal{\PRD}{54}{5395}{1996}.

\bibitem{CaloPerf}A. Airapetian {\it et al}, CERN-LHCC-96-40.

\bibitem{Dice}A. Artamonov {\it et al}, ATLAS-SOFT/95-14c.

\bibitem{Kawagoe:2003jv}K. Kawagoe {\it et al}, \Journal{\PRD}{69}{035003}{2004}.

\bibitem{Hinchliffe:1998ys}I. Hinchliffe and F.E. Paige, \Journal{\PRD}{60}{095002}{1999}.


\end{thebibliography}
\end{document}